\documentclass[aps,pra,notitlepage,showpacs,superscriptaddress,amsmath,amssymb]{revtex4-1}
\usepackage{graphicx}

\begin{document}

\title{Superluminal two-color light in multiple Raman gain medium}

\author{V. Kudria\v{s}ov}
\email{viaceslav.kudriasov@ff.vu.lt}
\affiliation{Institute of Theoretical Physics and Astronomy, Vilnius University,
A. Go\v{s}tauto 12, Vilnius 01108, Lithuania}

\author{J. Ruseckas}
\email{julius.ruseckas@tfai.vu.lt}
\affiliation{Institute of Theoretical Physics and Astronomy, Vilnius University,
A. Go\v{s}tauto 12, Vilnius 01108, Lithuania}

\author{A. Mekys}
\affiliation{Institute of Theoretical Physics and Astronomy, Vilnius University,
A. Go\v{s}tauto 12, Vilnius 01108, Lithuania}

\author{A. Ekers}
\altaffiliation[Present address: ]{King Abdullah University of Science and
  Technology, Thuwal 23955-6900, Kingdom of Saudi Arabia}
\affiliation{University of Latvia, Institute of Atomic Physics and Spectroscopy,
  LV-1586 Riga, Latvia}

\author{N. Bezuglov}
\affiliation{Faculty of Physics, St.\ Petersburg State University, 198904 St.\ Petersburg,
Russia}
\affiliation{University ITMO, Kronverkskiy pr.\ 49, St.\ Petersburg 197101,
  Russia}

\author{G. Juzeli\=unas}
\affiliation{Institute of Theoretical Physics and Astronomy, Vilnius University,
A. Go\v{s}tauto 12, Vilnius 01108, Lithuania}

\begin{abstract}
We investigate theoretically the formation of two-component light
with superluminal group velocity in a medium controlled by four Raman
pump fields. In such an optical scheme only a particular combination
of the probe fields is coupled to the matter and exhibits superluminal
propagation, the orthogonal combination is uncoupled. The individual probe fields
do not have a definite group velocity in the medium. Calculations demonstrate
that this superluminal component experiences an envelope advancement in
the medium with respect to the propagation in vacuum.
\end{abstract}

\pacs{42.50.Nn, 42.65.Hy, 42.25.Hz}

\maketitle

\section{Introduction}

The concepts of light velocity and speed of information transfer have
been debated by many outstanding scientists in the past \cite{Rayleigh-PM-1899,Brillouin-BOOK-1960,Born-Wolf-BOOK-1997}.
It is commonly accepted that the ultimate limitation for the speed
of information transfer is imposed by the causality principle. According
to it, no information can be transferred at the speed exceeding the
speed of light in vacuum $c$. Particularly, for light pulses it
means that the motion of the front of a light pulse or the energy
transport cannot occur at velocities greater than $c$ \cite{Brillouin-BOOK-1960,Ware-OE-2001,Stenner-Nature-2003}.

The phase and group velocities of a light pulse
have no such strict limitations. They may take arbitrary values depending
on the material properties and be significantly different from the vacuum
speed of light \cite{Milonni-BOOK-2005}. Group velocity defines the
speed of propagation of the envelope of the pulse
\begin{equation}
v_{g}=\frac{c}{n_{g}}=\frac{c}{n+\omega(dn/d\omega)}\,,
\end{equation}
where $n$ and $n_{g}$ are refractive and group velocity indices,
respectively. The group velocity can be managed through the dispersion
control of the medium $dn/d\omega$. Under the normal dispersion conditions
$dn/d\omega>0$, the group velocity is always less than the phase
velocity in the medium, $c/n$. It is possible to reach extremely
small values of $v_{g}$ (called ``slow light'') in the case of
electromagnetically induced transparency \cite{Arimondo1996,Harris1997,Scully1997,Lukin2003,Fleischhauer2005},
where steep dispersive profile over a short wavelength range is achieved
\cite{Hau-Nature-1999}. On the contrary, in the case of anomalous dispersion when
$dn/d\omega<0$, the group velocity becomes higher than $c$ if
\begin{equation}
n+\omega(dn/d\omega)<1\,.
\end{equation}
Moreover, the group velocity changes its sign to a negative value
in the case
\begin{equation}
n+\omega(dn/d\omega)<0\,.
\end{equation}
Both these conditions correspond to superluminal (fast) light propagation
regime. In this regime the pulse traverses the medium
at the speed exceeding that in the vacuum. 

There have been plenty of remarkable works devoted to the topic of
superluminal propagation (see e.g.\ \cite{Jiang-PRA-2007} and references
therein). The conditions of anomalous dispersion can be naturally achieved
within the medium's absorption band \cite{Chu-PRL-1982} or inside
a tunnel barrier \cite{Steinberg-PRL-1993}. Although possible, superluminal
propagation in this case is hardly observed due to significant loss
\cite{Chu-PRL-1982} or pulse deformations \cite{Chiao-ProgOptic-1997}.
To avoid that, some novel approaches have been suggested to use transparent
spectral regions for superluminal light \cite{Chiao-PRA-1993,Steinberg-PRA-1994,Dogariu-PRA-2001,Gehring-Science-2006,Glasser-PRL-2012}.
Particularly, it was demonstrated that in the case of Raman gain doublet
almost linear anomalous dispersion can be created and therefore distorsionless pulse
propagation is possible \cite{Dogariu-PRA-2001}.

Theoretical considerations predict rather striking and counterintuitive
features of superluminal light. Among the most exciting is a propagation
in backward direction (backward light) or appearance of pulse peak
at the exit of medium prior to entering (negative transit time) \cite{Chu-PRL-1982,Dogariu-PRA-2001}.
Recent experimental observations confirmed both of these predictions
\cite{Dogariu-PRA-2001,Gehring-Science-2006}. Although being apparently
extraordinary, these phenomena arise, in fact, due to the rephasing
of pulse spectral components favored by the anomalous dispersion \cite{Dogariu-PRA-2001}.
The associated energy transport always occurs in the forward direction
\cite{Gehring-Science-2006} and its speed is strictly limited by
the vacuum speed of light \cite{Ware-OE-2001}.

Most schemes employ only a single frequency probe pulse to produce
superluminal light or demonstrate simultaneous formation of slow and
fast light
\cite{Zhang-OL-2006,Bianucci-PRA-2008,Pati-OE-2009,Patnaik-OL-2011}.
The propagation of a single frequency probe beam in an N-type atomic
system using double Raman gain process is investigated in
Ref.~\cite{Bacha2013}. In this work we consider a different concept
where superluminal light is achieved for the superposition of probe
fields at different frequencies. Formation of such coupled optical
fields (spinor light) were first analyzed in
\cite{Unanyan-PRL-2010,Ruseckas-PRA-2011}. Here, we demonstrate that
similar spinor properties can also be achieved in the case of fast
light. In the following we analyze the propagation of two probe fields
amplified by Raman process pumped by four strong time-independent
fields. Such a scheme may be viewed as consisting of two Raman gain
doublets each providing the amplification for the corresponding spectral
components of the probe pulse. We begin with the well-known
Raman amplification schemes where single probe and either single or
double pump beams and used. Based on these results we derive the
propagation equations for the two probe fields in the presence of one
Raman gain doublet (two pump frequencies) and two Raman gain doublets
(four pump frequencies). Our study demonstrates that the latter scheme
supports the uncoupled and coupled states where the coupled state
exhibits superluminal propagation. Mathematically the coupled
state is represented as a linear superposition of the probe field
envelopes which has a definite group velocity (exceeding c).  In
this scheme the superluminal propagation is possible only for
specific superpositions of the field envelopes, and the individual
probe fields do not have definite group velocities. This is a novel
aspect of the wave mixing in Raman media which may be interesting for
optical signal control or interferometric applications.

The paper is organized as follows: In Section \ref{sec:equations}
we derive the equations of the propagation of the two monochromatic
probe fields in the presence of multiple Raman resonances. We use
the monochromatic solutions of the field equations further in Section
\ref{sec:propagation} to investigate the superluminal propagation
of the two-component wavepackets of light. Section \ref{sec:conclusions}
summarizes our findings and discusses some possible experimental implementations 
of the superluminal light.

\section{\label{sec:equations}Equations for the propagation of the probe
fields}

In this Section we derive the equations describing the propagation
of the probe field(s) in the Raman gain configuration. In order to
present the physical situation more transparently we start with a
simplest scheme containing only one probe field and one pump field.
After describing the propagation of the probe field in this simplest
scheme we consider more complicated schemes with additional pump and
probe fields.

\subsection{Single probe field}

\subsubsection{Single pump field}

\begin{figure}
\includegraphics[width=0.3\textwidth]{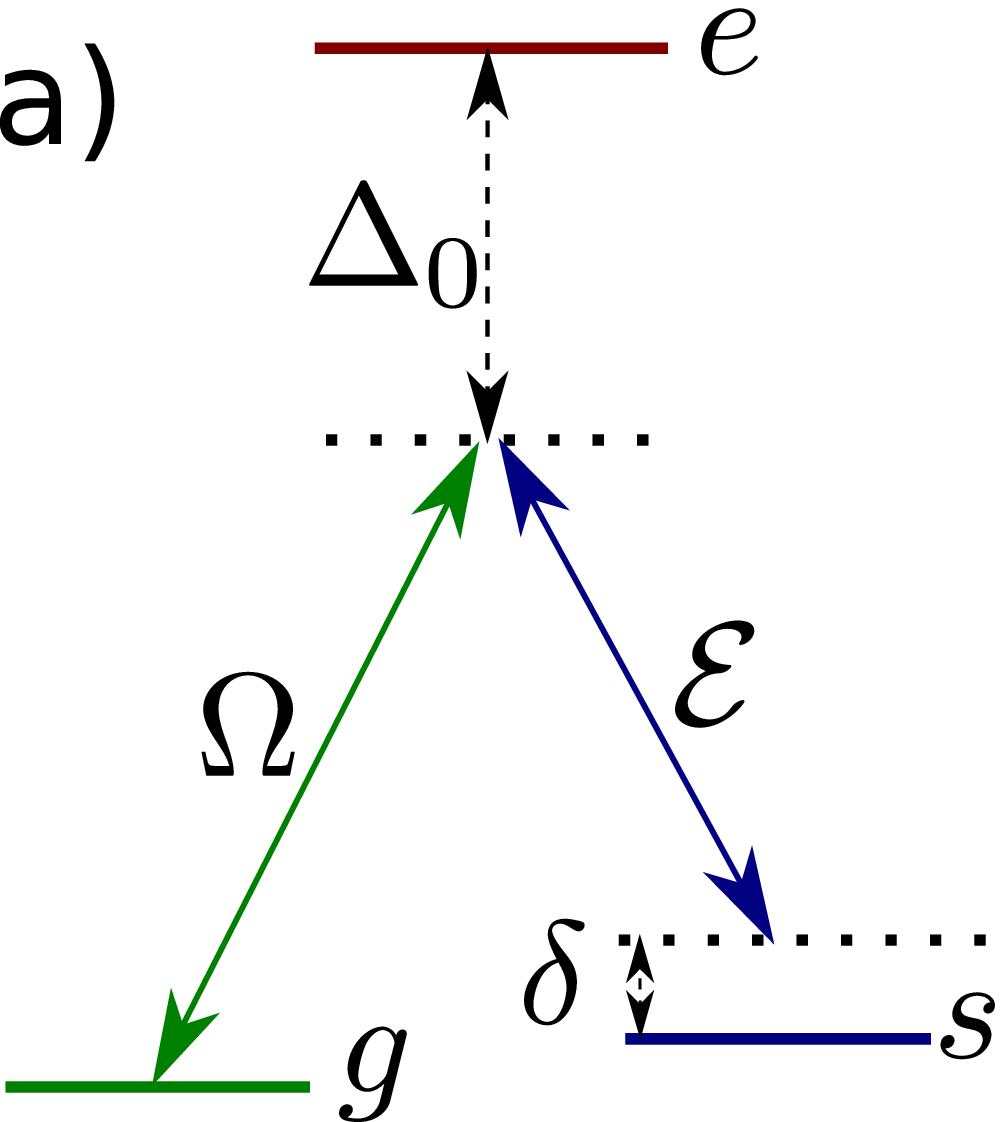}\hspace{1cm}\includegraphics[width=0.29\textwidth]{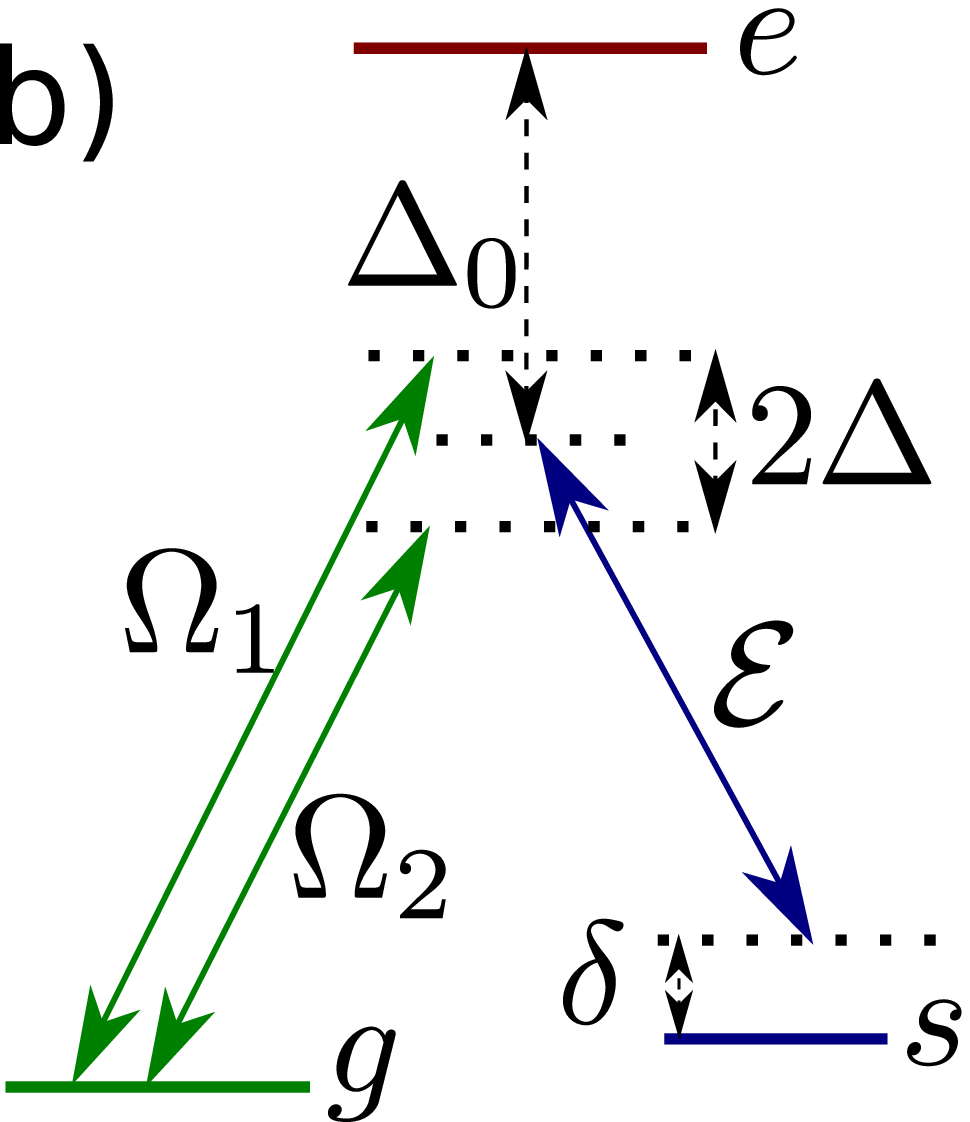}
\caption{(Color online) Raman amplification schemes with a single probe field: Raman singlet
(a) and Raman doublet (b)}
\label{fig:single-probe}
\end{figure}

First we consider the simplest Raman amplification scheme shown in
Fig.~\ref{fig:single-probe}a. Suppose there is an ensemble of atoms
characterized by two hyperfine ground levels $g$ and $s$, and an
electronic excited level $e$. The state of the atoms is described by
atomic amplitudes $\Psi_{g}(\mathbf{r},t)$, $\Psi_{s}(\mathbf{r},t)$,
$\Psi_{e}(\mathbf{r},t)$. The atoms interact with two light fields: a
strong pump laser and a weaker probe field. Initially the atoms are in
the ground level $g$ and we assume the Rabi frequency and duration of
the probe pulse are small enough so that we can neglect the depletion of
the ground level $g$. The propagation of the probe field inside of the
atomic cloud we describe similarly as in Ref.~\cite{Unanyan-PRL-2010}.
We write the electric field of the probe beam in the form of a plane
wave with modulated amplitude propagating along the $z$ axis:
\begin{equation}
\mathbf{E}(\mathbf{r},t)=\hat{\mathbf{e}}\sqrt{\frac{\hbar\omega}{2\varepsilon_{0}}}\mathcal{E}(\mathbf{r},t)e^{ikz-i\omega t}+\mathrm{H.c.}\,.\label{eq:probe field}
\end{equation}
Here $\omega$ is the central frequency of the probe beam, $k=\omega/c$
is the corresponding wave vector, and $\hat{\mathbf{e}}\bot\hat{\mathbf{z}}$
is the unit polarization vector. The probe field $\mathbf{E}(\mathbf{r},t)$
obeys the following wave equation:
\begin{equation}
c^{2}\nabla^{2}\mathbf{E}-\frac{\partial^{2}}{\partial t^{2}}\mathbf{E}=\frac{1}{\varepsilon_{0}}\frac{\partial^{2}}{\partial t^{2}}\mathbf{P}\,,\label{eq:wave equation}
\end{equation}
where
\begin{equation}
\mathbf{P}=\hat{\mathbf{e}}\mu\Psi_{s}^{*}\Psi_{e}+\mathrm{H.c.}\label{eq:polarization}
\end{equation}
is the polarization field of atoms, $\mu$ being the dipole moment
for the atomic transition $s\rightarrow e$. The atomic amplitudes
are normalized according to the equation $|\Psi_{g}|^{2}+|\Psi_{s}|^{2}+|\Psi_{e}|^{2}=n$,
where $n$ is the atomic density.  We introduce the slowly varying polarization $\mathcal{P}$ as:
\begin{equation}
\mathbf{P}=\hat{\mathbf{e}}\mathcal{P}e^{ikz-i\omega t}+\mathrm{H.c.}
\end{equation}
Using Eq.~(\ref{eq:polarization}) we get
\begin{equation}
\mathcal{P}=\mu\Psi_{s}^{*}\Psi_{e}e^{i\omega t-ikz}\,.\label{eq:slow-p}
\end{equation}
In case when
the amplitude $\mathcal{E}$ varies slowly during the wavelength and
optical cycle we can approximate Eq.~(\ref{eq:wave equation}) as \cite{Allen1975}
\begin{equation}
(\partial_{t}+c\partial_{z})\mathcal{E}=i\tilde{g}\mathcal{P}\,,
\end{equation}
where
\begin{equation}
\tilde{g}=\sqrt{\frac{\omega}{2\varepsilon_{0}\hbar}}\,.
\end{equation}
Let us introduce the slowly varying
atomic amplitudes
\begin{eqnarray}
\Phi_{g} & = & \Psi_{g}e^{i\omega_{g}t}\,,\\
\Phi_{s} & = & \Psi_{s}e^{i(\omega_{g}+\omega_{p}-\omega)t+ikz}\,,\\
\Phi_{e} & = & \Psi_{e}e^{i(\omega_{g}+\omega_{p})t}\,,
\end{eqnarray}
where $\hbar\omega_{g}$ is the energy of the atomic ground state $g$ and
$\omega_{p}$ is the frequency of the pump field. Using the slowly
varying atomic amplitudes and Eq.~(\ref{eq:slow-p}) the slowly varying
polarization $\mathcal{P}$ can be written as
\begin{equation}
\mathcal{P}=\mu\Phi_{s}^{*}\Phi_{e}
\end{equation}

The equations for the slowly varying atomic amplitudes are
\begin{eqnarray}
i\partial_{t}\Phi_{e} & = & \Delta_{0}\Phi_{e}-g\mathcal{E}\Phi_{s}-\Omega\Phi_{g}\,,\\
i\partial_{t}\Phi_{s} & = & (\delta-i\gamma)\Phi_{s}-g\mathcal{E}^{*}\Phi_{e}\,,
\end{eqnarray}
where $\Omega$ is the Rabi frequency of the pump field,
\begin{equation}
\Delta_{0}=\omega_{e}-\omega_{g}-\omega_{p}
\end{equation}
is one-photon detuning and
\begin{equation}
\delta=\omega-\omega_{p}+\omega_{s}-\omega_{g}
\end{equation}
is two-photon detuning. Here $\hbar\omega_{e}$ and $\hbar\omega_{s}$
are energies of the atomic states $e$ and $s$. The parameter $\gamma$
characterizes the decay rate of the level $s$ and the parameter $g=\mu\tilde{g}$ characterizes the strength of coupling of the probe field with the
atoms.

Let us consider monochromatic probe field for which the amplitude
$\mathcal{E}$ is time-independent. We search for time-independent
atomic amplitudes $\Phi_{g}$, $\Phi_{s}$ and $\Phi_{e}$, so that
\begin{eqnarray}
c\partial_{z}\mathcal{E}-ig\Phi_{s}^{*}\Phi_{e} & = & 0\label{eq:electric}\\
\Delta_{0}\Phi_{e}-g\mathcal{E}\Phi_{s}-\Omega\Phi_{g} & = & 0\label{eq:e}\\
(\delta-i\gamma)\Phi_{s}-g\mathcal{E}^{*}\Phi_{e} & = & 0\label{eq:s}
\end{eqnarray}
When one-photon detuning $\Delta_{0}$ is large, $\Delta_{0}|\delta-i\gamma|\gg g^{2}|\mathcal{E}|^{2}$
, Eq.~(\ref{eq:e}) yields
\begin{equation}
\Phi_{e}=\frac{\Omega}{\Delta_{0}}\Phi_{g}\,.\label{eq:e-g}
\end{equation}
Substituting Eq.~(\ref{eq:e-g}) into Eq.~(\ref{eq:s}) we get
\begin{equation}
\Phi_{s}=\frac{g\Phi_{g}\Omega}{\Delta_{0}(\delta-i\gamma)}\mathcal{E}^{*}\,.\label{eq:s-E}
\end{equation}
Finally, using Eqs.~(\ref{eq:e-g}), (\ref{eq:s-E}), and (\ref{eq:electric})
we get the propagation equation for the electric field
\begin{equation}
c\partial_{z}\mathcal{E}-i\frac{g^{2}n|\Omega|^{2}}{\Delta_{0}^{2}(\delta+i\gamma)}\mathcal{E}=0\,.\label{eq:electric-2}
\end{equation}
Here we have taken into account that $|\Phi_{g}|^{2}=n$ in accordance with the adopted normalization.
Plane waves
\begin{equation}
\mathcal{E}(z)=\mathcal{E}(0)e^{i\kappa z}
\end{equation}
with 
\begin{equation}
\kappa=\frac{g^{2}n|\Omega|^{2}}{c\Delta_{0}^{2}(\delta+i\gamma)}\,.\label{eq:kappa}
\end{equation}
are the solutions of Eq.~(\ref{eq:electric-2}). Note, that $\kappa$
depends on the frequency $\omega$ of the probe field via the two-photon
detuning $\delta$. The group velocity of the probe field in the medium
can be determined from $\kappa(\omega)$. Since the fast-varying amplitude
is proportional to $E(z,t)\sim\mathcal{E}(0)\exp\{i(\omega/c+\kappa(\omega))z-i\omega t\}$,
the maximum of the wave packet made from such plane waves moves with
the group velocity
\begin{equation}
v_{g}=c\left\{ \frac{d(\omega+c\textrm{Re}\kappa(\omega))}{d\omega}\right\} ^{-1}=\frac{c}{1+\frac{\partial}{\partial\omega}\textrm{Re}\left\{ c\kappa(\omega)\right\} }\,.
\end{equation}
Using Eq.~(\ref{eq:kappa}) we get
\begin{equation}
v_{g}=\frac{c}{1+\frac{g^{2}n|\Omega|^{2}}{\Delta_{0}^{2}}\frac{\gamma^{2}-\delta^{2}}{(\delta^{2}+\gamma^{2})^{2}}}\,.\label{eq:group velocity}
\end{equation}
One can see that in the case $\gamma<\delta$ the group velocity exceeds
$c$. This situation corresponds to the wings of the gain profile
where the dispersion is anomalous. However, when two-photon detuning
$\delta$ is close to zero, $v_{g}<c$. In order to improve this situation
and have group velocity larger than $c$ in Ref.~\cite{Steinberg-PRA-1994}
it was suggested to use two pump fields with different frequencies.

The amplitude of the probe field propagating through atomic cloud
is changed. If the monochromatic probe field is incident on the atomic
cloud, the amplitude of the transmitted field at the end of the atomic
cloud becomes 
\begin{equation}
\mathcal{E}(L)\sim\exp(i\kappa L)=\exp\left(i\frac{g^{2}n|\Omega|^{2}L}{c\Delta_{0}^{2}(\delta+i\gamma)}\right)\,.
\end{equation}
By separating real and imaginary parts we obtain the transmission
coefficient
\begin{equation}
R=\exp\left(\frac{|\Omega|^{2}}{\Delta_{0}^{2}}\frac{\gamma g^{2}nL}{c(\delta^{2}+\gamma^{2})}\right)=\exp\left(\frac{|\Omega|^{2}}{\Delta_{0}^{2}}\frac{L}{L_{\mathrm{dec}}}\frac{\gamma^{2}}{\delta^{2}+\gamma^{2}}\right)\,,
\end{equation}
where
\begin{equation}
L_{\mathrm{dec}}=\frac{\gamma c}{g^{2}n}\label{eq:ldec}
\end{equation}
is the characteristic length related to the decay of the level $s$.
Since the expression in the exponent is positive, the transmission
coefficient $R>1$, so there is an amplification of the probe beam.

\subsubsection{Two pump fields (Raman doublet)}

Now let us consider a situation where two strong pump beams (with
frequencies $\omega_{p_{1}}$ and $\omega_{p_{2}}$ ) act on the atomic
ensemble instead of one pump beam. This situation corresponds to a
Raman gain doublet (Fig.~\ref{fig:single-probe}b) and was investigated
in Refs.~\cite{Steinberg-PRA-1994,Wang2000,Dogariu-PRA-2001,Kuzmich2001}.
The consistent mathematical description of this case can be obtained
using Floquet theory \cite{Chu2004}. However, here we make use of
a simpler approach. To describe the propagation of the probe beam
in this scheme we separate the atomic amplitudes into two parts oscillating
with different frequencies: $\Psi_{e}=\Psi_{e_{1}}+\Psi_{e_{2}}$,
$\Psi_{s}=\Psi_{s_{1}}+\Psi_{s_{2}}$ with corresponding slowly changing
amplitudes
\begin{eqnarray*}
\Phi_{e_{1}} & = & \Psi_{e_{1}}e^{i(\omega_{g}+\omega_{p_{1}})t}\,,\qquad\Phi_{e_{2}}=\Psi_{e_{2}}e^{i(\omega_{g}+\omega_{p_{2}})t}\,,\\
\Phi_{s_{1}} & = & \Psi_{s_{1}}e^{i(\omega_{g}+\omega_{p_{1}}-\omega)t+ikz}\,,\qquad\Phi_{s_{2}}=\Psi_{s_{2}}e^{i(\omega_{g}+\omega_{p_{2}}-\omega)t+ikz}\,.
\end{eqnarray*}
After separating the atomic amplitudes into two parts,
Eq.~(\ref{eq:slow-p}) yields the following relation for the slowly
varying polarization:
\begin{equation}
\mathcal{P}=\mu(\Phi_{s_{1}}^{*}\Phi_{e_{1}}+\Phi_{s_{2}}^{*}\Phi_{e_{2}}
+\Phi_{s_{1}}^{*}\Phi_{e_{2}}e^{-2i\Delta t}+\Phi_{s_{2}}^{*}\Phi_{e_{1}}e^{2i\Delta t})\,,
\end{equation}
where
\begin{equation}
2\Delta=\omega_{p_{2}}-\omega_{p_{1}}\,.
\end{equation}
Neglecting the terms oscillating with a large  frequency $2\Delta$ which is still small compared to
the one-photon detuning $\Delta_0$,
one can write equations for the probe field and atomic amplitudes
as 
\begin{eqnarray}
(\partial_{t}+c\partial_{z})\mathcal{E} & = & ig\Phi_{s_{1}}^{*}\Phi_{e_{1}}+ig\Phi_{s_{2}}^{*}\Phi_{e_{2}}\,,\\
i\partial_{t}\Phi_{e_{1}} & = & (\Delta_{0}-\Delta)\Phi_{e_{1}}-g\mathcal{E}\Phi_{s_{1}}-\Omega_{1}\Phi_{g}\,,\\
i\partial_{t}\Phi_{e_{2}} & = & (\Delta_{0}+\Delta)\Phi_{e_{2}}-g\mathcal{E}\Phi_{s_{2}}-\Omega_{2}\Phi_{g}\,,\\
i\partial_{t}\Phi_{s_{1}} & = & (\delta+\Delta-i\gamma)\Phi_{s_{1}}-g\mathcal{E}^{*}\Phi_{e_{1}}\,,\\
i\partial_{t}\Phi_{s_{2}} & = & (\delta-\Delta-i\gamma)\Phi_{s_{2}}-g\mathcal{E}^{*}\Phi_{e_{2}}\,.
\end{eqnarray}
Here
\begin{equation}
\Delta_{0}=\omega_{e}-\omega_{g}-\frac{1}{2}(\omega_{p_{1}}+\omega_{p_{2}})
\end{equation}
is an average one-photon detuning and
\begin{equation}
\delta=\omega-\frac{1}{2}(\omega_{p_{1}}+\omega_{p_{2}})+\omega_{s}-\omega_{g}\label{eq:detun-2}
\end{equation}
is an average two-photon detuning. Proceeding similarly as in the
case of one pump field we obtain the following set of equations for
the time-independent complex amplitudes in the case of large $\Delta_{0}$:
\begin{eqnarray}
c\partial_{z}\mathcal{E} & = & ig\Phi_{s_{1}}^{*}\frac{\Omega_{1}}{\Delta_{0}}\Phi_{g}+ig\Phi_{s_{2}}^{*}\frac{\Omega_{1}}{\Delta_{0}}\Phi_{g}\,,\label{eq:35}\\
\Phi_{s_{1}} & = & \frac{g\Omega_{1}}{\Delta_{0}(\delta+\Delta-i\gamma)}\Phi_{g}\mathcal{E}^{*}\,,\\
\Phi_{s_{2}} & = & \frac{g\Omega_{2}}{\Delta_{0}(\delta-\Delta-i\gamma)}\Phi_{g}\mathcal{E}^{*}\,.\label{eq:37}
\end{eqnarray}
From Eqs.~(\ref{eq:35})--(\ref{eq:37}) follows the equation for
the probe field 
\begin{equation}
c\partial_{z}\mathcal{E}-i\frac{g^{2}n}{\Delta_{0}^{2}}\left[\frac{|\Omega_{1}|^{2}}{\delta+\Delta+i\gamma}+\frac{|\Omega_{2}|^{2}}{\delta-\Delta+i\gamma}\right]\mathcal{E}=0\,.\label{eq:electric-3}
\end{equation}
Searching for the plane wave solution we find
\begin{equation}
\kappa=\frac{g^{2}n}{c\Delta_{0}^{2}}\left[\frac{|\Omega_{1}|^{2}}{\delta+\Delta+i\gamma}+\frac{|\Omega_{2}|^{2}}{\delta-\Delta+i\gamma}\right]\,.\label{eq:doublet kappa}
\end{equation}
In a particular case where $|\Omega_{1}|=|\Omega_{2}|$ and $\delta=0$
one finds
\begin{equation}
\frac{\partial}{\partial\omega}\textrm{Re}\left\{ c\kappa(\omega)\right\} =-2g^{2}n\frac{|\Omega_{1}|^{2}}{\Delta_{0}^{2}}\frac{\Delta^{2}-\gamma^{2}}{(\Delta^{2}+\gamma^{2})^{2}}\,.
\end{equation}
Thus for $\Delta>\gamma$ the group velocity is larger than $c$.
We see that, in contrast to the scheme with a single pump field, we
have superluminal propagation even for zero two-photon detuning $\delta$.

\subsection{Two probe fields}

\subsubsection{Two pump fields}

\begin{figure}
\includegraphics[width=0.3\textwidth]{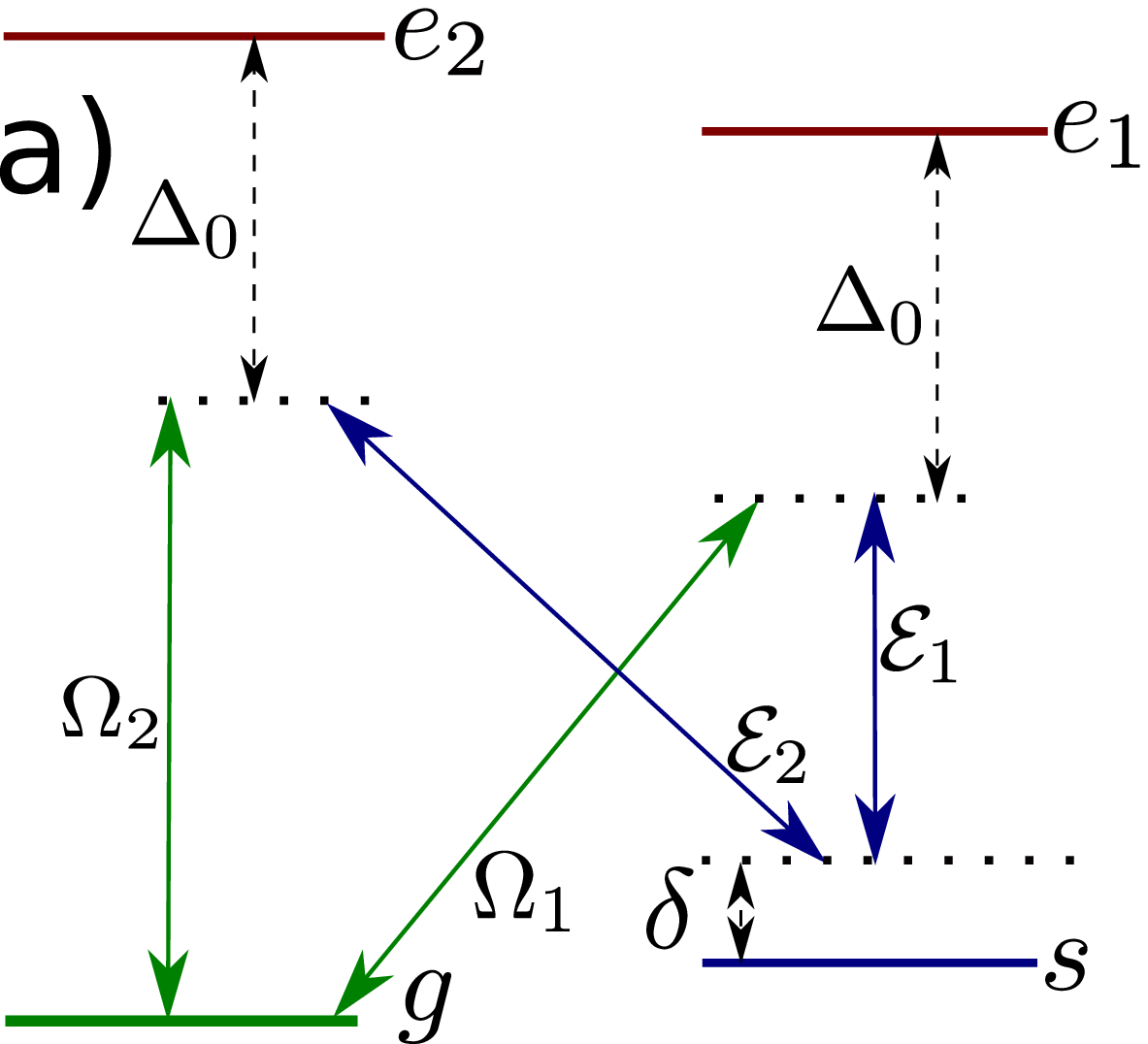}\hspace{1cm}\includegraphics[width=0.3\textwidth]{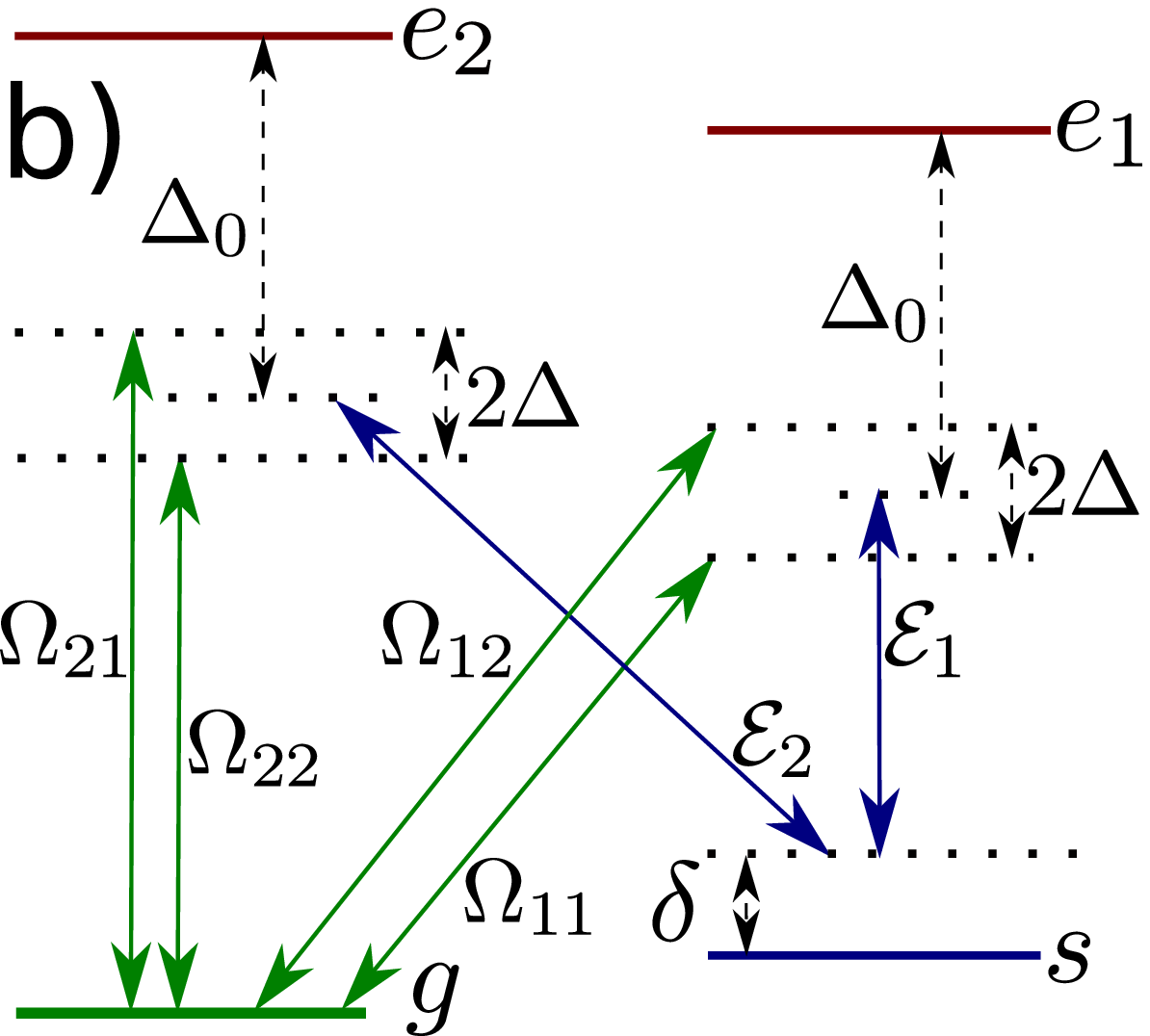}
\caption{(Color online) Raman amplification schemes with two probe fields: double Raman singlet
(a) and double Raman doublet (b)}
\label{fig:double-probe}
\end{figure}

Now we shall turn to the main goal of this article by studying the
propagation of two probe fields in Raman gain situation. Thus we consider
an ensemble of atoms characterized by two hyperfine ground levels
$g$, $s$ and two electronic excited levels $e_{1}$ and $e_{2}$.
The state of the atoms is described by the atomic amplitudes $\Psi_{g}(\mathbf{r},t)$,
$\Psi_{s}(\mathbf{r},t)$, $\Psi_{e_{1}}(\mathbf{r},t)$, $\Psi_{e_{2}}(\mathbf{r},t)$.
Similarly to the single probe field we first investigate a simpler
scheme where the atoms interact with four light fields: two strong
pump lasers and two weaker probe fields (Fig.~\ref{fig:double-probe}a).
We assume the four photon resonance condition
\begin{equation}
\omega_{1}-\omega_{p_{1}}=\omega_{2}-\omega_{p_{2}}\,,
\end{equation}
where $\omega_{1}$ and $\omega_{2}$ are frequencies of the probe
beams, $\omega_{p_{1}}$ and $\omega_{p_{2}}$ are frequencies of
the pump beams.

For each probe beam we introduce slowly varying amplitudes $\mathcal{E}_{1}$
and $\mathcal{E}_{2}$ of the electric field according to Eq.~(\ref{eq:probe field}).
Wave vectors of the probe fields are $k_{1}=\omega_{1}/c$ and $k_{2}=\omega_{2}/c$
. The wave equations and the corresponding polarization fields are
written separately for each of the probe field similarly to Eqs.~(\ref{eq:wave equation})
and (\ref{eq:polarization}). In the following, the strength of coupling
of probe fields with the atoms is assumed to be the same for both
probe fields: $g=\mu_{1}\sqrt{\omega_{1}/2\varepsilon_{0}\hbar}=\mu_{2}\sqrt{\omega_{2}/2\varepsilon_{0}\hbar}\,,$where
$\mu_{1}$ and $\mu_{2}$ denote the dipole momenta for the atomic
transitions $s\rightarrow e_{1}$ and $s\rightarrow e_{2}$, respectively.
After introducing the slowly varying atomic amplitudes we obtain the
following equations for slowly varying probe field amplitudes $\mathcal{E}_{1}$
and $\mathcal{E}_{2}$ :
\begin{eqnarray}
(\partial_{t}+c\partial_{z})\mathcal{E}_{1} & = & ig\Phi_{s}^{*}\Phi_{e_{1}}\,,\\
(\partial_{t}+c\partial_{z})\mathcal{E}_{2} & = & ig\Phi_{s}^{*}\Phi_{e_{2}}\,.
\end{eqnarray}
On the other hand, the equations for the atomic amplitudes are
\begin{eqnarray}
i\partial_{t}\Phi_{e_{1}} & = & \Delta_{0}\Phi_{e_{1}}-g\mathcal{E}_{1}\Phi_{s}-\Omega_{1}\Phi_{g}\,,\\
i\partial_{t}\Phi_{e_{2}} & = & \Delta_{0}\Phi_{e_{2}}-g\mathcal{E}_{2}\Phi_{s}-\Omega_{2}\Phi_{g}\,,\\
i\partial_{t}\Phi_{s} & = & (\delta-i\gamma)\Phi_{s}-g\mathcal{E}_{1}^{*}\Phi_{e_{1}}-g\mathcal{E}_{2}^{*}\Phi_{e_{2}}\,,
\end{eqnarray}
where
\begin{equation}
\Delta_{0}=\omega_{e_{1}}-\omega_{g}-\omega_{p_{1}}=\omega_{e_{2}}-\omega_{g}-\omega_{p_{2}}
\end{equation}
is one-photon detuning and
\begin{equation}
\delta=\omega_{1}-\omega_{p_{1}}+\omega_{s}-\omega_{g}=\omega_{2}-\omega_{p_{2}}+\omega_{s}-\omega_{g}
\end{equation}
is two-photon detuning. Here $\hbar\omega_{e_{1}}$, $\hbar\omega_{e_{2}}$
and $\hbar\omega_{s}$ are energies of the atomic states $e_{1}$,
$e_{2}$ and $s$, respectively.

As before, we consider the case of monochromatic probe beams with
time independent amplitudes $\mathcal{E}_{1}$, $\mathcal{E}_{2}$
and the constant atomic amplitudes $\Phi_{g}$, $\Phi_{s}$, $\Phi_{e_{1}}$
and $\Phi_{e_{2}}$. Assuming a large detuning, $\Delta_{0}|\delta-i\gamma|\gg g^{2}|\mathcal{E}|^{2}$,
the corresponding equations for the atomic amplitudes reduce to 
\begin{eqnarray}
\Phi_{e_{1}} & = & \frac{\Omega_{1}}{\Delta_{0}}\Phi_{g}\,,\label{eq:e1-g}\\
\Phi_{e_{2}} & = & \frac{\Omega_{2}}{\Delta_{0}}\Phi_{g}\,.\label{eq:e2-g}
\end{eqnarray}
and 
\begin{equation}
\Phi_{s}=\frac{g\Phi_{g}}{\Delta_{0}(\delta-i\gamma)}(\Omega_{1}\mathcal{E}_{1}^{*}+\Omega_{2}\mathcal{E}_{2}^{*})\,.\label{eq:s-E-1}
\end{equation}
Substituting these relations into the equations for the fields $\mathcal{E}_{1}$
and $\mathcal{E}_{2}$, we get
\begin{eqnarray}
c\partial_{z}\mathcal{E}_{1}-i\frac{g^{2}n\Omega_{1}}{\Delta_{0}^{2}(\delta+i\gamma)}(\Omega_{1}^{*}\mathcal{E}_{1}+\Omega_{2}^{*}\mathcal{E}_{2}) & = & 0\,,\label{eq:electric1-2}\\
c\partial_{z}\mathcal{E}_{2}-i\frac{g^{2}n\Omega_{2}}{\Delta_{0}^{2}(\delta+i\gamma)}(\Omega_{1}^{*}\mathcal{E}_{1}+\Omega_{2}^{*}\mathcal{E}_{2}) & = & 0\,.\label{eq:electric2-2}
\end{eqnarray}

Introducing new fields representing superpositions of the original
probe fields
\begin{eqnarray}
\psi & = & \frac{1}{\Omega}(\Omega_{1}^{*}\mathcal{E}_{1}+\Omega_{2}^{*}\mathcal{E}_{2})\,,\\
\phi & = & \frac{1}{\Omega}(\Omega_{2}\mathcal{E}_{1}-\Omega_{1}\mathcal{E}_{2})\,,
\end{eqnarray}
Eqs.~(\ref{eq:electric1-2}) and (\ref{eq:electric2-2}) take the
form
\begin{eqnarray}
c\partial_{z}\psi-i\frac{g^{2}n\Omega^{2}}{\Delta_{0}^{2}(\delta+i\gamma)}\psi & = & 0\,,\label{eq:psi}\\
c\partial_{z}\phi & = & 0\,,
\end{eqnarray}
where
\begin{equation}
\Omega=\sqrt{|\Omega_{1}|^{2}+|\Omega_{2}|^{2}}
\end{equation}
is the total Rabi frequency. One can see that the field $\phi$ propagates
like in free space without interacting with the atoms. The other field
$\psi$ does interact with the atoms. The solutions of Eq.~(\ref{eq:psi})
are plane waves:
\begin{equation}
\psi(z)=\psi(0)e^{i\kappa z}\,,
\end{equation}
with
\begin{equation}
\kappa=\frac{g^{2}n\Omega^{2}}{c\Delta_{0}^{2}(\delta+i\gamma)}\,.\label{eq:kappa-1}
\end{equation}
This result coincides with the Eq.~(\ref{eq:kappa}), implying that
the group velocity has the form of Eq.~(\ref{eq:group velocity}).
For $\gamma<\delta$ the group velocity exceeds the vacuum speed of
light. 

\subsubsection{Four pump fields (double Raman doublet)}

Let us now consider a situation where four strong pump beams act on
the atomic ensemble. This situation corresponds to a Raman gain doublet
for each of the probe beams (Fig.~\ref{fig:double-probe}b). We assume
four-photon resonances
\begin{eqnarray}
\omega_{p_{1,1}}-\omega_{1} & = & \omega_{p_{2,1}}-\omega_{2}\,,\\
\omega_{p_{1,2}}-\omega_{1} & = & \omega_{p_{2,2}}-\omega_{2}\,,
\end{eqnarray}
where $\omega_{p_{1,1}}$, $\omega_{p_{1,2}}$, $\omega_{p_{2,1}}$
and $\omega_{p_{2,2}}$ are the frequencies of the pump beams. Similarly
as in the scheme with the single probe beam we write the atomic amplitudes
as a sum of two parts: $\Psi_{e_{1}}=\Psi_{e_{1,1}}+\Psi_{e_{1,2}}$,
$\Psi_{e_{2}}=\Psi_{e_{2,1}}+\Psi_{e_{2,2}}$, $\Psi_{s}=\Psi_{s_{1}}+\Psi_{s_{2}}$.
Introducing the slowly changing amplitudes and neglecting the terms
oscillating with the frequency
\begin{equation}
2\Delta=\omega_{p_{1,2}}-\omega_{p_{1,1}}=\omega_{p_{2,2}}-\omega_{p_{2,1}}
\end{equation}
we find the following set of equations
\begin{eqnarray}
c\partial_{z}\mathcal{E}_{1} & = & ig\Phi_{s_{1}}^{*}\frac{\Omega_{1,1}}{\Delta_{0}}\Phi_{g}+ig\Phi_{s_{2}}^{*}\frac{\Omega_{1,2}}{\Delta_{0}}\Phi_{g}\label{eq:65}\\
c\partial_{z}\mathcal{E}_{2} & = & ig\Phi_{s_{1}}^{*}\frac{\Omega_{2,1}}{\Delta_{0}}\Phi_{g}+ig\Phi_{s_{2}}^{*}\frac{\Omega_{2,2}}{\Delta_{0}}\Phi_{g}\\
\Phi_{s_{1}} & = & \frac{g\Phi_{g}}{(\delta+\Delta-i\gamma)\Delta_{0}}(\Omega_{1,1}\mathcal{E}_{1}^{*}+\Omega_{2,1}\mathcal{E}_{2}^{*})\\
\Phi_{s_{2}} & = & \frac{g\Phi_{g}}{(\delta-\Delta-i\gamma)\Delta_{0}}(\Omega_{1,2}\mathcal{E}_{1}^{*}+\Omega_{2,2}\mathcal{E}_{2}^{*})\label{eq:68}
\end{eqnarray}
for the amplitudes of the monochromatic probe fields and the time
independent atomic amplitudes. Here
\begin{equation}
\Delta_{0}=\omega_{e_{1}}-\omega_{g}-\frac{1}{2}(\omega_{p_{1,1}}+\omega_{p_{1,2}})=\omega_{e_{2}}-\omega_{g}-\frac{1}{2}(\omega_{p_{2,1}}+\omega_{p_{2,2}})
\end{equation}
is an average two-photon detuning and
\begin{equation}
\delta=\omega_{1}-\frac{1}{2}(\omega_{p_{1,1}}+\omega_{p_{1,2}})+\omega_{s}-\omega_{g}=\omega_{2}-\frac{1}{2}(\omega_{p_{2,1}}+\omega_{p_{2,2}})+\omega_{s}-\omega_{g}
\end{equation}
is an average two-photon detuning. From Eqs.~(\ref{eq:65})--(\ref{eq:68})
we obtain the equations of propagation of the probe fields
\begin{eqnarray}
c\partial_{z}\mathcal{E}_{1} & = & i\frac{g^{2}n}{\Delta_{0}^{2}}\left[\frac{\Omega_{1,1}(\Omega_{1,1}^{*}\mathcal{E}_{1}+\Omega_{2,1}^{*}\mathcal{E}_{2})}{(\delta+\Delta+i\gamma)}+\frac{\Omega_{1,2}(\Omega_{1,2}^{*}\mathcal{E}_{1}+\Omega_{2,2}^{*}\mathcal{E}_{2})}{(\delta-\Delta+i\gamma)}\right]\,,\\
c\partial_{z}\mathcal{E}_{2} & = & i\frac{g^{2}n}{\Delta_{0}^{2}}\left[\frac{\Omega_{2,1}(\Omega_{1,1}^{*}\mathcal{E}_{1}+\Omega_{2,1}^{*}\mathcal{E}_{2})}{(\delta+\Delta+i\gamma)}+\frac{\Omega_{2,2}(\Omega_{1,2}^{*}\mathcal{E}_{1}+\Omega_{2,2}^{*}\mathcal{E}_{2})}{(\delta-\Delta+i\gamma)}\right]\,.
\end{eqnarray}

Let us consider a particular situation in which
\begin{equation}
\frac{\Omega_{1,2}}{\Omega_{1,1}}=\frac{\Omega_{2,2}}{\Omega_{2,1}}\,.
\end{equation}
Introducing new fields
\begin{eqnarray}
\psi & = & \frac{1}{\Omega_{1}}(\Omega_{1,1}^{*}\mathcal{E}_{1}+\Omega_{2,1}^{*}\mathcal{E}_{2})\label{eq:field-psi}\\
\phi & = & \frac{1}{\Omega_{1}}(\Omega_{2,1}\mathcal{E}_{1}-\Omega_{1,1}\mathcal{E}_{2})\label{eq:field-fi}
\end{eqnarray}
we get the equations for the fields $\psi$ and $\phi$: 
\begin{eqnarray}
c\partial_{z}\psi-i\frac{g^{2}n}{\Delta_{0}^{2}}\left[\frac{\Omega_{1}^{2}}{\delta+\Delta+i\gamma}+\frac{\Omega_{2}^{2}}{\delta-\Delta+i\gamma}\right]\psi & = & 0\,,\label{eq:psi-3}\\
c\partial_{z}\phi & = & 0\,,
\end{eqnarray}
where
\begin{eqnarray}
\Omega_{1} & = & \sqrt{|\Omega_{1,1}|^{2}+|\Omega_{2,1}|^{2}}\\
\Omega_{2} & = & \sqrt{|\Omega_{1,2}|^{2}+|\Omega_{2,2}|^{2}}
\end{eqnarray}
The field $\phi$ propagates without interaction with atoms. The plane
wave solution of Eq.~(\ref{eq:psi-3}) gives
\begin{equation}
\kappa=\frac{g^{2}n}{c\Delta_{0}^{2}}\left[\frac{|\Omega_{1}|^{2}}{\delta+\Delta+i\gamma}+\frac{|\Omega_{2}|^{2}}{\delta-\Delta+i\gamma}\right]\,.\label{eq:kappa4}
\end{equation}
This quantity corresponds to Eq.~(\ref{eq:doublet kappa}) providing
a superluminal propagation.

\section{\label{sec:propagation}Propagation of probe beam wave packets}

To illustrate the superluminal behavior of the probe pulses, in this
Section we will consider the propagation of a Gaussian wave packet
through the atomic cloud. The wave packet is formed by taking a superposition
of monochromatic solutions of the propagation equations. The length
of the atomic cloud is $L$. For simplicity we will measure all frequencies
in units of $\gamma$ and time in units of $\gamma^{-1}$ and set
$\gamma=1$. Furthermore, by measuring the length in the units of
$c/\gamma$, we set $c=1$.

\subsection{Single probe field}

At first we will consider the propagation of the incident Gaussian
wave packet for a scheme with a single probe beam and a Raman gain
doublet, as shown in Fig.~\ref{fig:single-probe}b. The fast-varying
amplitude of the monochromatic probe field is
\begin{equation}
E_{\delta}(z,t)=\begin{cases}
\mathcal{E}_{\delta}e^{i\delta z-i\delta t}\,, & z\leqslant0\\
\mathcal{E}_{\delta}e^{i(\delta+\kappa(\delta))z-i\delta t}\,, & 0<z<L\\
\mathcal{E}_{\delta}e^{i\kappa(\delta)L+i\delta z-i\delta t}\,, & L\leqslant z\,.
\end{cases}\label{eq:E_delta}
\end{equation}
Here we have used the two-photon detuning $\delta$ (\ref{eq:detun-2})
instead of the frequency $\omega$ . The change of the wave number
$\kappa(\delta)$ is given by Eq.~(\ref{eq:doublet kappa}). When
$\Omega_{1}=\Omega_{2}$ we can write 
\begin{equation}
\kappa=M\left[\frac{1}{\delta+\Delta+i}+\frac{1}{\delta-\Delta+i}\right]\,,\label{eq:kappa-2}
\end{equation}
with
\begin{equation}
M=\frac{1}{L_{\mathrm{dec}}}\frac{|\Omega_{1}|^{2}}{\Delta_{0}^{2}}\,.
\end{equation}
Here $L_{\mathrm{dec}}$ is the length defined by Eq.~(\ref{eq:ldec}).
At the central frequency $\delta=0$ the group velocity is
\begin{equation}
v_{g}=\frac{1}{1+\kappa^{\prime}(0)}=\frac{1}{1-2M\frac{\Delta^{2}-1}{(\Delta^{2}+1)^{2}}}\,.
\end{equation}
In order to get the group velocity larger than $1$ (i.e. larger than
$c=1$), the dimensionless one-photon detuning should be $\Delta>1$.
The group velocity is maximum when $\Delta=\sqrt{3}$. For this value
of $\Delta$ it is $v_{g}=(1-M/4)^{-1}$. If $M>4$, we have a negative
group velocity. For $\delta=0$ the transmission coefficient is
\begin{equation}
R=\exp(i\kappa(0)L)=\exp\left(ML\frac{2}{\Delta^{2}+1}\right)\,.\label{eq:R-1}
\end{equation}
In particular, for $\Delta=\sqrt{3}$ , the transmission coefficient
is $R=\exp(ML/2)$.

The Gaussian wave packet can be formed by taking a superposition of
monochromatic waves (\ref{eq:E_delta})
\begin{equation}
E(z,t)=\int_{-\infty}^{+\infty}E_{\delta}(z,t)\, d\delta\,,\label{eq:E_int}
\end{equation}
with
\begin{equation}
\mathcal{E}_{\delta}=\frac{1}{\sqrt{\pi}\sigma}\exp\left(-\frac{\delta^{2}}{\sigma^{2}}-i\delta z_{0}\right)\,,\label{eq:e-delta}
\end{equation}
where $z_{0}$ is a location of the initial wavepacket peak, and $\sigma$
is the width of the packet in the frequency domain. To be in the dispersion
region with a negative slope, the width $\sigma$ should be smaller
than approximately $0.8$. Using Eq.~(\ref{eq:e-delta}) the incident
probe field reads
\begin{equation}
E(z,t)=\exp\left[-\frac{\sigma^{2}}{4}(z-z_{0}-t)^{2}\right]\,.\label{eq:gauss}
\end{equation}
From this equation we see that the width of the wave packet in the
coordinate space is
\begin{equation}
\sigma_{z}=\frac{2}{\sigma}\,.
\end{equation}
To avoid tails of the initial wave packet in the atomic cloud, we
need to have $|z_{0}|\gg\sigma_{z}$.

For a Gaussian packet narrow in the frequency space, we can obtain
approximate expressions for the electric field by expanding $\kappa(\delta)$
in power series and taking the first three terms: 
\begin{equation}
\kappa(\delta)\approx\kappa(0)+\kappa^{\prime}(0)\delta+\kappa^{\prime\prime}\delta^{2}/2\,,\label{eq:kappa-delta}
\end{equation}
with
\begin{eqnarray}
\kappa(0) & = & -2i\frac{M}{\Delta^{2}+1}\\
\kappa^{\prime}(0) & = & -2M\frac{\Delta^{2}-1}{(\Delta^{2}+1)^{2}}\\
\kappa^{\prime\prime}(0) & = & -4iM\frac{3\Delta^{2}-1}{(\Delta^{2}+1)^{3}}
\end{eqnarray}
Nonlinear terms in the expansion are associated with group velocity
dispersion and cause pulse distortion. After performing the integration
we get approximate expressions for the probe field. The probe field
is
\begin{equation}
E_{\mathrm{inside}}(z,t)\approx\frac{1}{\sqrt{1-i\frac{1}{2}\kappa^{\prime\prime}(0)\sigma^{2}z}}\exp\left(-\frac{\sigma^{2}[(1+\kappa^{\prime}(0))z-z_{0}-t]^{2}}{4\left(1-i\frac{1}{2}\kappa^{\prime\prime}(0)\sigma^{2}z\right)}+i\kappa(0)z\right)\label{eq:e-inside}
\end{equation}
inside the atomic medium and
\begin{equation}
E_{\mathrm{outside}}(z,t)\approx\frac{1}{\sqrt{1-i\frac{1}{2}\kappa^{\prime\prime}(0)\sigma^{2}L}}\exp\left(-\frac{\sigma^{2}[z+\kappa^{\prime}(0)L-z_{0}-t]^{2}}{4\left(1-i\frac{1}{2}\kappa^{\prime\prime}(0)\sigma^{2}L\right)}+i\kappa(0)L\right)\label{eq:e-outside}
\end{equation}
outside the atomic cloud. From Eq.~(\ref{eq:e-outside}) it follows
that the distortion of the pulse is determined by the parameter $[1-i\kappa^{\prime\prime}(0)\sigma^{2}L/2]^{-1/2}$
\cite{Dogariu-PRA-2001}.

\begin{figure}
\includegraphics[width=0.35\textwidth]{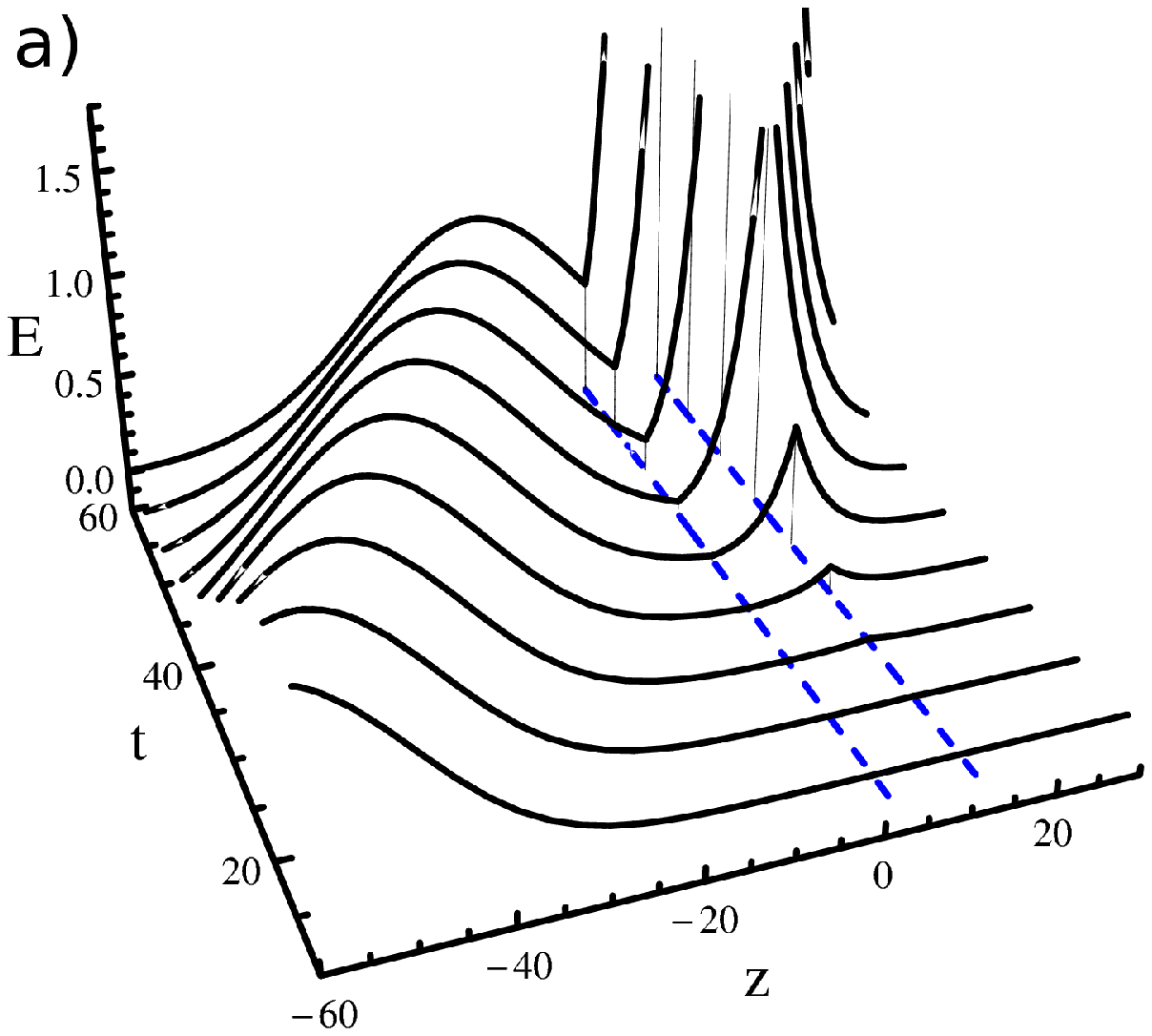}\includegraphics[width=0.45\textwidth]{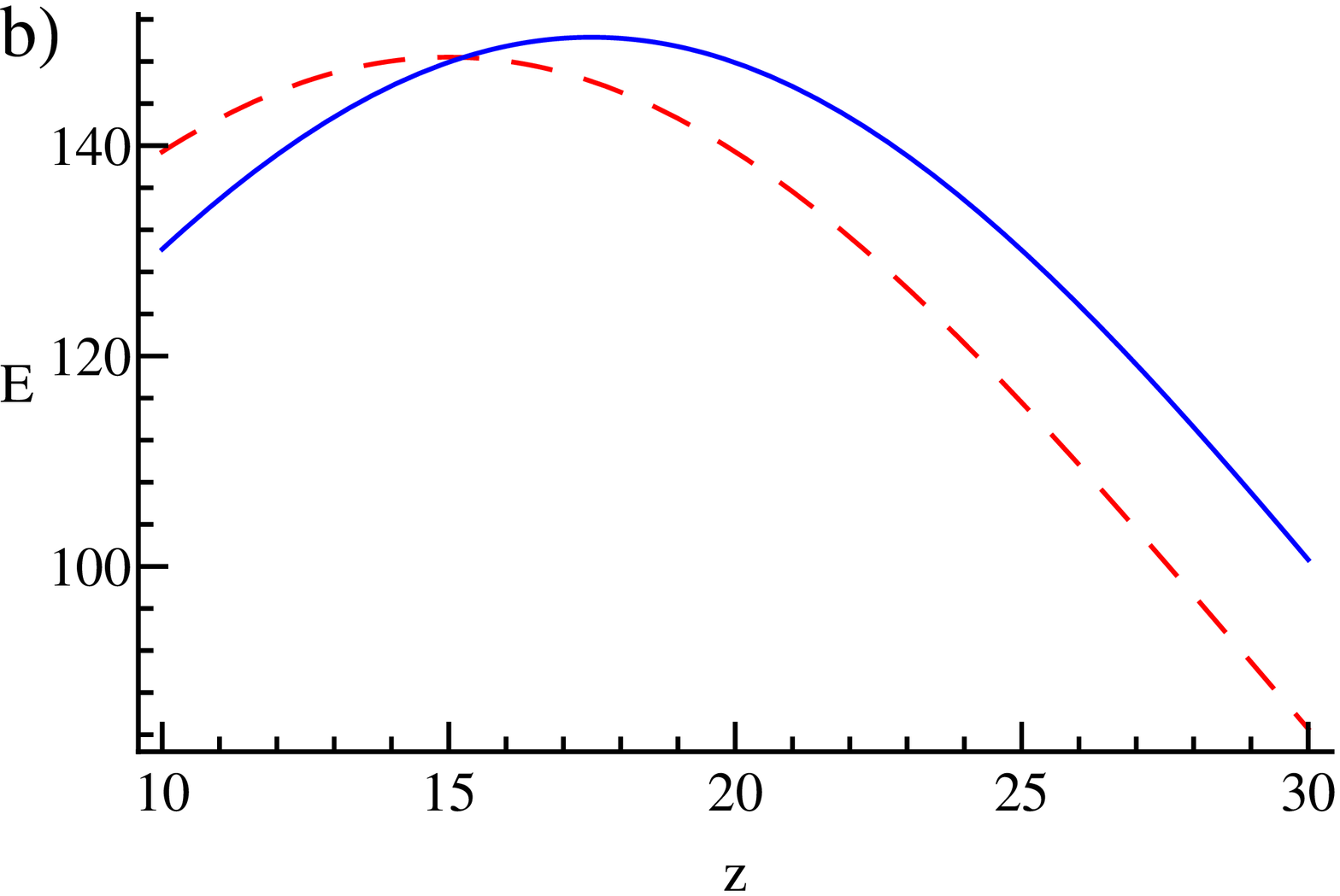}
\caption{(Color online) Propagation of the incoming Gaussian wave packet, described by Eq.~(\ref{eq:gauss})
through atomic cloud, calculated using Eqs.~(\ref{eq:E_delta}),
(\ref{eq:E_int}), (\ref{eq:e-delta}). All the quantities shown are
dimensionless. (a) Time evolution of the probe field. Dashed blue
lines show the location of the atomic cloud. (b) Comparison of the
transmitted wave packet (solid blue line) with the wave packet propagating
in the vacuum (dashed red line) at the same time moment $t=90$. In
order to make the amplitudes of the wave packets similar, the amplitude
of the Gaussian packet propagating in the vacuum is increased by the
factor $R$ given by Eq.~(\ref{eq:R-1}). The parameters used in
calculation are $\sigma=0.1$, $z_{0}=-75$, $\Delta=\sqrt{3}$, $M=1$;
the length of the atomic cloud is $L=10$. At these parameters the
group velocity is $v_{g}=1.33$ and transmission coefficient $R=148.4$.}
\label{fig:packet-single}
\end{figure}

Propagation of the Gaussian wave packet, described by Eq.~(\ref{eq:gauss}),
through atomic cloud is shown in Fig.~\ref{fig:packet-single}. As
one can see in Fig.~\ref{fig:packet-single}a, the front tail of
the wave packet, entering the atomic cloud gets amplified and develops
a maximum at the other end of the atomic cloud. Comparison of the
transmitted wave packet with the wave packet propagating in the vacuum
at the same time moment is shown in Fig.~\ref{fig:packet-single}b.
In order to make the amplitudes of the wave packets similar, the amplitude
of the wave packet propagating in the vacuum is increased by the factor
$R$ given by Eq.~(\ref{eq:R-1}). One can see that the maximum of
the wave packet after the atomic cloud is located at larger value
of the coordinate $z$ than the maximum of the wave packet propagating
in the vacuum. This is a signature of the superluminal group velocity,
$v_{g}>1$.

\subsection{Two probe fields}

Next let us consider the propagation of the incident Gaussian wave
packet for the scheme with four pump beams (two Raman gain doublets),
shown in Fig.~\ref{fig:double-probe}b. At first we will consider
propagation of monochromatic probe fields. Let us assume that only
one probe field $\mathcal{E}_{1}$ is incident on the atomic cloud.
The amplitude of this probe field at the beginning of the atomic cloud
is $\mathcal{E}_{1}(z=0)\equiv\mathcal{E}_{\delta}$. Here we use
the two-photon detuning 
\begin{equation}
\delta=\omega_{1}-\frac{1}{2}(\omega_{p_{1,1}}+\omega_{p_{1,2}})+\omega_{s}-\omega_{g}=\omega_{2}-\frac{1}{2}(\omega_{p_{2,1}}+\omega_{p_{2,2}})+\omega_{s}-\omega_{g}\,.
\end{equation}
instead of the frequencies $\omega_{1}$ and $\omega_{2}$. The fields
$\psi$ and $\phi$, introduced by Eqs.~(\ref{eq:field-psi}), (\ref{eq:field-fi}),
at the beginning of the atomic cloud are 
\begin{eqnarray}
\psi(z=0) & = & \frac{\Omega_{1,1}^{*}}{\Omega}\mathcal{E}_{\delta}\,,\\
\phi(z=0) & = & \frac{\Omega_{2,1}}{\Omega}\mathcal{E}_{\delta}\,.
\end{eqnarray}
Inside the atomic cloud the fields $\psi$ and $\phi$ depend on the
coordinate $z$ according to $\phi(z)=\phi(0)$ and $\psi(z)=\psi(0)e^{i\kappa z}$,
with $\kappa$ given by Eq.~(\ref{eq:kappa4}). Thus, at the end
of the cloud the fields $\psi$ and $\phi$ are $\phi(L)=\phi(0)$
and $\psi(L)=e^{i\kappa L}\psi(0)$ . We will consider only the case
when $\Omega_{1}=\Omega_{2}$. Then the expression for the wave number
$\kappa(\delta)$ is the same as in the scheme with the single probe
beam and is given by Eq.~(\ref{eq:kappa-2}). The electric fields
of the probe beams inside the atomic cloud can be obtained from the
fields $\psi$ and $\phi$: 
\begin{eqnarray}
\mathcal{E}_{1}(z) & = & \frac{1}{\Omega_{1}}(\Omega_{1,1}\psi(z)+\Omega_{2,1}^{*}\phi(z))=\left(1+\frac{|\Omega_{1,1}|^{2}}{\Omega_{1}^{2}}(e^{i\kappa z}-1)\right)\mathcal{E}_{\delta}\,,\label{eq:tmp-e1}\\
\mathcal{E}_{2}(z) & = & \frac{1}{\Omega_{1}}(\Omega_{2,1}\psi(z)-\Omega_{1,1}^{*}\phi(z))=\frac{\Omega_{2,1}\Omega_{1,1}^{*}}{\Omega_{1}^{2}}(e^{i\kappa z}-1)\mathcal{E}_{\delta}\,.\label{eq:tmp-e2}
\end{eqnarray}
Using Eqs.~(\ref{eq:tmp-e1}), (\ref{eq:tmp-e2}), the fast-varying
amplitude of the monochromatic probe fields are
\begin{eqnarray}
E_{1,\delta}(z,t) & = & \begin{cases}
\mathcal{E}_{\delta}e^{i\delta z-i\delta t}\,, & z\leqslant0\,,\\
\left(1+\frac{|\Omega_{1,1}|^{2}}{\Omega_{1}^{2}}(e^{i\kappa(\delta)z}-1)\right)\mathcal{E}_{\delta}e^{i\delta z-i\delta t}\,, & 0<z<L\,,\\
\left(1+\frac{|\Omega_{1,1}|^{2}}{\Omega_{1}^{2}}(e^{i\kappa(\delta)L}-1)\right)\mathcal{E}_{\delta}e^{i\delta z-i\delta t}\,, & L\leqslant z\,,
\end{cases}\label{eq:E1_delta}\\
E_{2,\delta}(z,t) & = & \begin{cases}
0\,, & z\leqslant0\,,\\
\frac{\Omega_{2,1}\Omega_{1,1}^{*}}{\Omega_{1}^{2}}(e^{i\kappa(\delta)z}-1)\mathcal{E}_{\delta}e^{i\delta z-i\delta t}\,, & 0<z<L\,,\\
\frac{\Omega_{2,1}\Omega_{1,1}^{*}}{\Omega_{1}^{2}}(e^{i\kappa(\delta)L}-1)\mathcal{E}_{\delta}e^{i\delta z-i\delta t}\,, & L\leqslant z\,.
\end{cases}\label{eq:E2_delta}
\end{eqnarray}
The amplitude of the second probe field $\mathcal{E}_{2}$ at the
other side of the atomic cloud is maximal when $|\Omega_{1,1}|/\Omega_{1}=|\Omega_{2,1}|/\Omega_{1}=1/\sqrt{2}$.

The Gaussian wave packet can be formed by taking superpositions of
monochromatic waves (\ref{eq:E1_delta}), (\ref{eq:E2_delta})
\begin{eqnarray}
E_{1}(z,t) & = & \int_{-\infty}^{+\infty}E_{1,\delta}(z,t)\, d\delta\label{eq:E1_int}\\
E_{2}(z,t) & = & \int_{-\infty}^{+\infty}E_{2,\delta}(z,t)\, d\delta\label{eq:E2_int}
\end{eqnarray}
with $\mathcal{E}_{\delta}$ given by Eq.~(\ref{eq:e-delta}). In
this case the electric field of the first probe beam in the free space
before the atomic cloud is given by Eq.~(\ref{eq:gauss}). After
performing the integration we obtain
\begin{eqnarray}
E_{1}(z,t) & = & \frac{|\Omega_{2,1}|^{2}}{\Omega_{1}^{2}}\exp\left(-\frac{\sigma^{2}}{4}(z-z_{0}-t)^{2}\right)+\frac{|\Omega_{1,1}|^{2}}{\Omega_{1}^{2}}E_{\mathrm{inside}}(z,t)\\
E_{2}(z,t) & = & \frac{\Omega_{2,1}\Omega_{1,1}^{*}}{\Omega_{1}^{2}}\left[E_{\mathrm{inside}}(z,t)-\exp\left(-\frac{\sigma^{2}}{4}(z-z_{0}-t)^{2}\right)\right]
\end{eqnarray}
for the probe fields inside of the atomic cloud and
\begin{eqnarray}
E_{1}(z,t) & = & \frac{|\Omega_{2,1}|^{2}}{\Omega_{1}^{2}}\exp\left(-\frac{\sigma^{2}}{4}(z-z_{0}-t)^{2}\right)+\frac{|\Omega_{1,1}|^{2}}{\Omega_{1}^{2}}E_{\mathrm{outside}}(z,t)\\
E_{2}(z,t) & = & \frac{\Omega_{2,1}\Omega_{1,1}^{*}}{\Omega_{1}^{2}}\left[E_{\mathrm{outside}}(z,t)-\exp\left(-\frac{\sigma^{2}}{4}(z-z_{0}-t)^{2}\right)\right]
\end{eqnarray}
for the probe fields at the other side of the atomic cloud. Here $E_{\mathrm{inside}}(z,t)$
and $E_{\mathrm{outside}}(z,t)$ are, respectively, the probe field
inside of the atomic cloud and after passing the atomic cloud in the
scheme with the single Raman gain doublet. For the incident Gaussian
packet narrow in the frequency space we can obtain approximate expressions
for the electric fields by expanding $\kappa(\delta)$ in power series.
Then $E_{\mathrm{inside}}(z,t)$ and $E_{\mathrm{outside}}(z,t)$
are given by Eqs.~(\ref{eq:e-inside}) and (\ref{eq:e-outside}).

\begin{figure}
\includegraphics[width=0.35\textwidth]{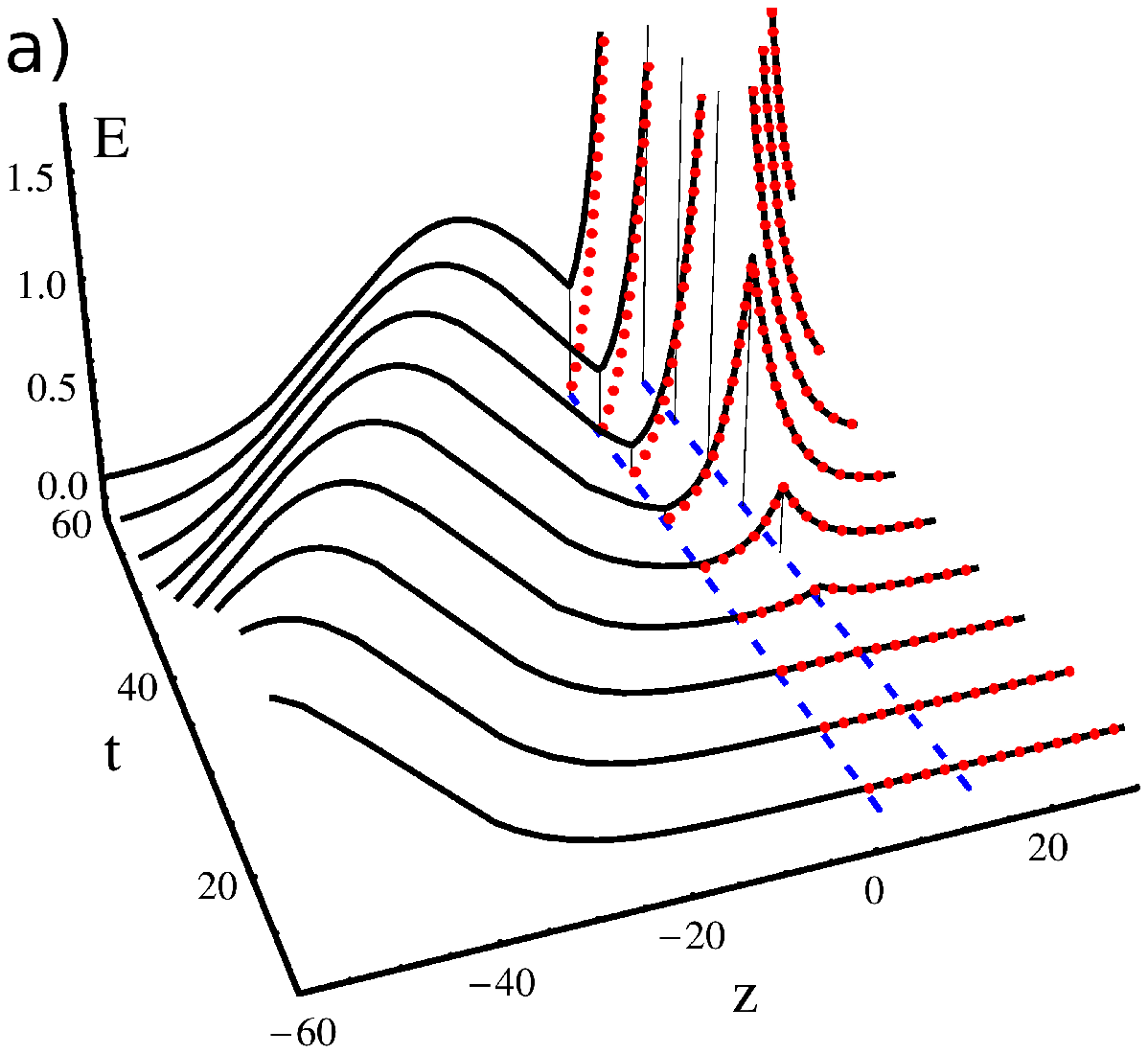}\includegraphics[width=0.45\textwidth]{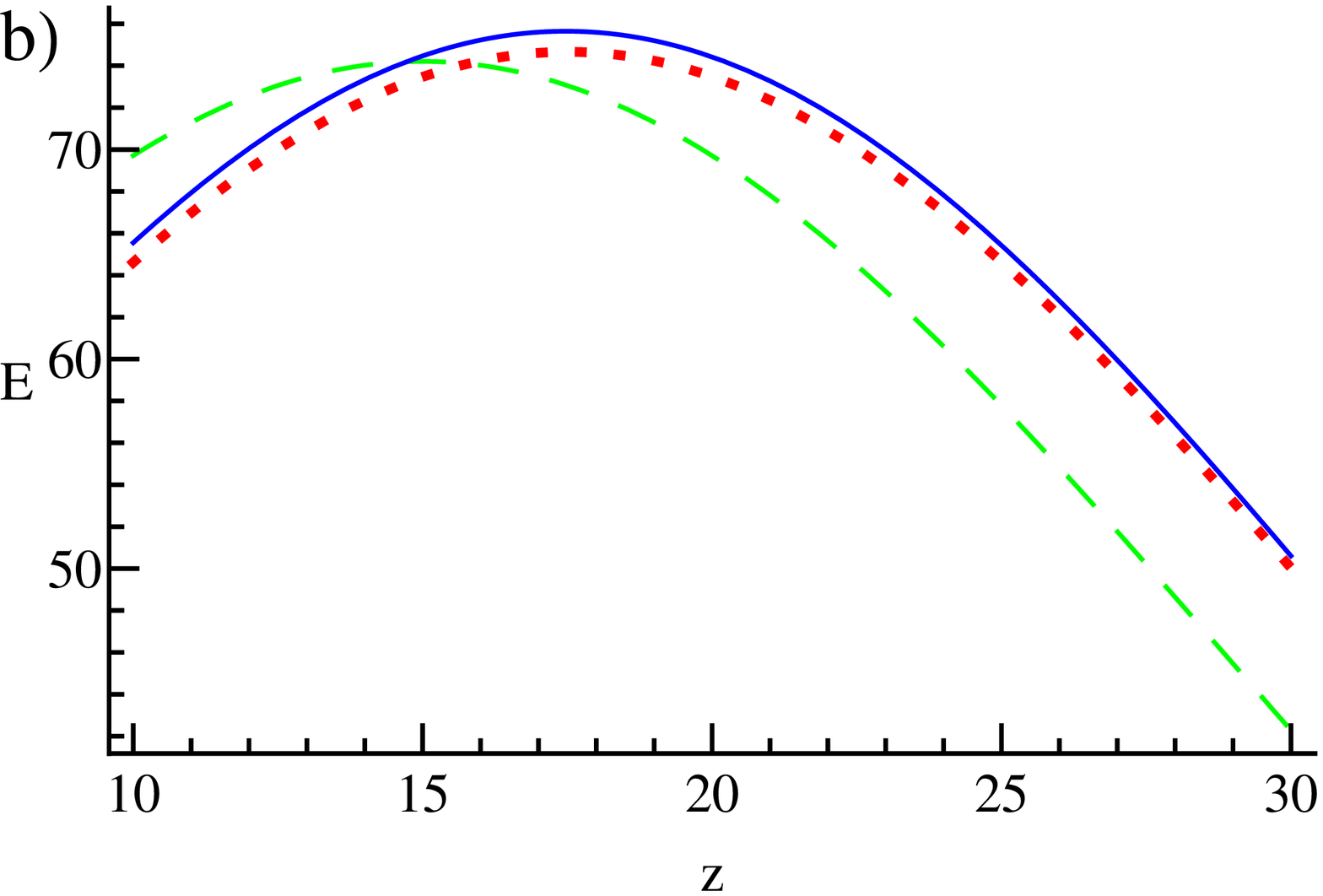}
\caption{(Color online) Propagation through the atomic cloud of the incoming Gaussian wave
packet (\ref{eq:gauss}) in the first probe field $\mathcal{E}_{1}$
and the creation of the second probe field $\mathcal{E}_{2}$ for
the scheme with two Raman gain doublets, calculated using Eqs.~(\ref{eq:E1_delta}),
(\ref{eq:E2_delta}), (\ref{eq:E1_int}), (\ref{eq:E2_int}), (\ref{eq:e-delta}).
All the quantities shown are dimensionless. (a) Evolution of the probe
fields with time. Solid black line shows the first probe field, dotted
red line shows the second probe field. Dashed blue lines indicate
the location of the atomic cloud. (b) Comparison of the wave packet
of the first probe beam (solid blue line) and the second probe beam
(dotted red line) with the incident wave packet of the first beam
propagating in the vacuum (dashed green line) at the same time moment
$t=90$. In order to make the amplitudes of the wave packets similar,
the amplitude of the Gaussian packet propagating in the vacuum is
increased by the factor $0.5R$, with $R$ given by Eq.~(\ref{eq:R-1}).
The parameters used in calculation are $\Omega_{1,1}/\Omega_{1}=\Omega_{2,1}/\Omega_{1}=1/\sqrt{2}$,
all other prameters are the same as in Fig.~\ref{fig:packet-single}.}
\label{fig:packet-double}
\end{figure}

The Fig.~\ref{fig:packet-double} illustrates the evolution of the
probe fields in the scheme with two Raman gain doublets when the first
probe field representing the Gaussian wave packet (\ref{eq:gauss})
is incident on the atomic cloud. As one can see in Fig.~\ref{fig:packet-double}a,
the front tail of the wave packet, entering the atomic cloud gets
amplified and develops a maximum at the other end of the atomic cloud.
In addition, the second probe field is created and also gets amplified,
reaching the maximum at the other end of the atomic cloud. The Fig.~\ref{fig:packet-double}b
compares the wave packets of the probe beams after exiting the atomic
cloud with the incoming wave packet of the first probe beam propagating
in the vacuum. In order to make the amplitudes of the wave packets
similar, the amplitude of the wave packet propagating in the vacuum
is increased by the factor $0.5R$ given by Eq.~(\ref{eq:R-1}).
The factor $0.5$ is needed because the energy in the scheme with
two Raman gain doublets is transferred to two probe beams, instead
of one beam in the scheme with single Raman gain doublet. After exiting
the atomic cloud, the maximum of the wave packet of the first probe
beam is seen to be located at larger value of the coordinate $z$
than the maximum of the wave packet propagating in the vacuum, indicating
the superluminal group velocity $v_{g}>1$. The maximum of the second
probe beam after the atomic cloud is almost at the same location as
the maximum of the first probe beam.

\section{\label{sec:conclusions}Concluding remarks}

We have demonstrated a possibility of producing superluminal light
composed of two probe waves characterized by different frequencies and
propagating in a medium with two Raman gain doublets. Although
individual probe fields exhibit Raman gain, a strong connection is
established between two probe fields due to the resonance between real
and virtual states in the coupling scheme. This leads to the formation
of a specific combination (superposition) of the probe field envelopes
propagating with a definite group velocity determined by the pump power
and the detunings. Such a regime corresponds to the pulse propagation
with a superluminal velocity and mathematically is described by a
particular solution of the wave equation. It is shown that a peak of the
superluminal wavepacket is advanced with respect to the corresponding
pulse propagating in the vacuum. Additionally, it is demonstrated that
if only one probe field is incident on the medium, both frequencies are
produced at the end of the medium as a result of the coupling between
the individual probe fields. Two-frequency superluminal light extends
possibilities to control light pulses and their interactions in optical
media.

The scheme for creating a two-component superluminal light, shown in
Fig.~\ref{fig:double-probe}b, can be experimentally implemented using an atomic
cesium (Cs) vapor cell at the room temperature, as in the experiment by Wang
\textit{et all.} \cite{Wang2000} on the single-component superluminal light. All
cesium atoms are to be prepared in the ground-state hyper-fine magnetic sublevel
$6\mathrm{S}_{1/2}$, $|F=4,m=-4\rangle$ serving as the level $g$ in our scheme.
The magnetic sublevel $6\mathrm{S}_{1/2}$, $|F=4,m=-2\rangle$ corresponds to the
level $s$. On the other hand, the states $6\mathrm{P}_{3/2}$, $|F=4,m=-3\rangle$
and $6\mathrm{P}_{1/2}$, $|F=4,m=-3\rangle$ can be chosen to be the excited
levels $e_1$ and $e_2$, respectively. The strong Raman pump beams should be
right-hand polarized ($\sigma^{+}$) and two weak Raman probe beams should be
left-hand polarized ($\sigma^{-}$) to couple properly the atomic levels. To
create the two-component supeluminal light, one can also make use of other
atoms, such as the rubidium ${}^{87}\mathrm{Rb}$ with the following hyper-fine magnetic
sublevels involved: $5\mathrm{S}_{1/2}$, $|F=2,m=-2\rangle$ as the ground level
$g$, $5\mathrm{S}_{1/2}$, $|F=2,m=0\rangle$ as the level $s$,
$5\mathrm{P}_{1/2}$, $|F=2,m=-1\rangle$ and $5\mathrm{P}_{3/2}$,
$|F=2,m=-1\rangle$ as the excited levels $e_1$ and $e_2$.

\section*{Acknowledgements}

This work has been supported by the project TAP LLT 01/2012 of the
Research Council of Lithuania, the National Science Council of Taiwan,
as well as the EU FP7 IRSES project COLIMA (Contract No. PIRSES-GA-2009-247475)
and the EU FP7 Centre of Excellence FOTONIKA-LV (REGPOT-CT-2011-285912-FOTONIKA).
N.B. acknowledges the partial support by Government of Russian Federation, Grant No. 074-U01.

\end{document}